\documentclass{article}

\usepackage{arxiv}

\usepackage[utf8]{inputenc} 
\usepackage[T1]{fontenc}    
\usepackage{hyperref}       
\usepackage{url}            
\usepackage{booktabs}       
\usepackage{amsfonts}       
\usepackage{nicefrac}       
\usepackage{microtype}      

\usepackage{graphicx}
\usepackage{algorithm}
\usepackage{algpseudocode}
\usepackage{svg}
\usepackage{amsmath}
\usepackage{hyperref}
\usepackage{amssymb}
\usepackage[backend=biber,style=chem-acs, eprint=false]{biblatex}
\author{Nicholas T. Runcie \\
	EaSTCHEM School of Chemistry\\
	University of Edinburgh\\
	Edinburgh, EH9 3FJ \\
	\texttt{N.Runcie@sms.ed.ac.uk}\
	\And Antonia S.J.S. Mey \\
	EaSTCHEM School of Chemistry\\
	University of Edinburgh\\
	Edinburgh, EH9 3FJ \\
	\texttt{antonia.mey@ed.ac.uk} \
}

\title{SILVR: Guided Diffusion for Molecule Generation}

\begin{document}

\maketitle

\begin{abstract}
Computationally generating novel synthetically accessible compounds with high affinity and low toxicity is a great challenge in drug design. Machine-learning models beyond conventional pharmacophoric methods have shown promise in generating novel small molecule compounds, but require significant tuning for a specific protein target. Here, we introduce a method called selective iterative latent variable refinement (SILVR) for conditioning an existing diffusion-based equivariant generative model without retraining. The model allows the generation of new molecules that fit into a binding site of a protein based on fragment hits. We use the SARS-CoV-2 Main protease fragments from Diamond X-Chem that form part of the COVID Moonshot project as a reference dataset for conditioning the molecule generation. The SILVR rate controls the extent of conditioning and we show that moderate SILVR rates make it possible to generate new molecules of similar shape to the original fragments, meaning that the new molecules fit the binding site without knowledge of the protein. We can also merge up to 3 fragments into a new molecule without affecting the quality of molecules generated by the underlying generative model. Our method is generalizable to any protein target with known fragments and any diffusion-based model for molecule generation. 
\end{abstract}

\section{Introduction}
\label{sec:intro}
Sampling from a very large space of possible drug-like compounds to find suitable hits for a given target protein is an open challenge in drug design. It is estimated that there are between $10^{23}$ to $10^{60}$ feasible compounds, while only around $10^8$ have been synthesised so far~\cite{polishchuk2013estimation, reymond2012exploring}. Different strategies have been used   to try and sample a diverse and synthetically accessible molecular space from pharmacophore search~\cite{schwab2010conformations} to machine learning-based methods. In particular, methods based on machine learning (ML) have shown vast promise in this space in recent years~\cite{bilodeau2022generative}. Various neural network architectures have been proposed for molecular generation, from variational autoencoders (VAE)~\cite{kingma2022autoencoding, jin2018junction, ma2018constrained, ragoza2022generating}, to generative adversarial networks (GAN)~\cite{hoffmann2019generating} and normalising flows~\cite{shi2021learning}. More recently denoising diffusion probabilistic models, and particularly equivariant diffusion models, have shown promise in molecular generation~\cite{xu2022geodiff, hoogeboom2022equivariant}. All of these were conceived primarily to generate new molecules, however, being able to generate chemically varied molecules is only the first hurdle for identifying new drug candidates.

Typically the objective is to generate a diverse set of molecules for a given target protein that are easily synthetically accessible, and ideally with high binding affinity and low predicted toxicity~\cite{prieto-martinez2019chapter}. A plethora of methods have been devised for the generation of molecules, as well as assessing their suitability as drug candidates. For example, for binding affinity predictions traditional docking~\cite{fu2018predictive} and molecular simulation-based affinity prediction methods~\cite{mey2016blinded,mey2020best} have dominated the field until recently. Now ML methods are gaining momentum and various approaches have been used to generate molecules for a binding site where in each case the training is conditioned towards the target protein~\cite{xie2022advances}. Some of these models incorporate a ligand score directly~\cite{ragoza2022generating}, while others require methods based on machine learning (ML)~\cite{masters2023deep}, or more conventional affinity prediction methods downstream (e.g. docking or free energy calculations). Even with a variety of ways to assess the synthetic accessibility of generated compounds~\cite{coley2018scscore, thakkar2021retrosynthetic}, molecules generated with these methods are often not easily synthesizable and in the worst cases can be chemically infeasible. 

Fragment-based drug discovery is an approach where a library of small molecular fragments ($<$300 Da) is screened against a target~\cite{hajduk2007decade, kumar2012fragment, bian2018computational}. These fragments are selected such that they present promiscuous binding, allowing exploration of many types of interactions a drug-like molecule could adopt within a given target. Individual fragments can not be drugs in and of themselves as they do not possess enough intermolecular interactions to achieve a sufficient binding affinity with a target, however, by considering an ensemble of known fragment hits, new high-affinity binders can be constructed by merging and linking known fragments together and elaborating on singular fragments. An array of screening methods exist for determining if fragments bind a target, however here we focus on X-ray crystallography techniques. Protein drug targets can either be co-crystallized with fragments or be crystalised unbound and subsequently soaked in a fragment solution. These crystals can be resolved by X-ray crystallography, with the results showing a high-quality electron density map revealing the exact binding geometry and interactions a fragment obtains with the given target. 
An application of generative models is therefore in the interpretation of such fragment data for the automated design of high-affinity binders. Goa \textit{et al.} have introduced a way to generate linkers for fragments using reinforcement learning strategies~\cite{guo2023linkinvent}, while Imrie \textit{et al.}~\cite{imrie2020deep} have used variational autoencoders on this task without the use of protein information. More recently Huang \textit{et al.}~\cite{huang20223dlinker} and Igashov \textit{et al.}~\cite{igashov2022equivariant} have tackled this challenge using equivariant variational auto-encoder and diffusion-based models respectively. Each of these models were explicitly trained for the specific purpose of linker generation.  

In this paper, we present SILVR, a selective iterative latent variable refinement (SILVR) method for conditioning an existing pre-trained equivariant diffusion model (EDM) towards the task of fragment merging and linker generation, yielding compounds similar to existing hits without specific training towards this task. To achieve this we combine the EDM by Hoogeboom \textit{et al.}~\cite{hoogeboom2022equivariant} with the iterative latent variable refinement method proposed by Choi \textit{et al.}~\cite{choi2021ilvr} for image generation networks. This allows the generation of new molecules in the shape of a binding site using information from existing fragment hits, without specific training or knowledge of this task. 

\begin{figure}[h!]
    \centering
    \includegraphics[width=\textwidth]{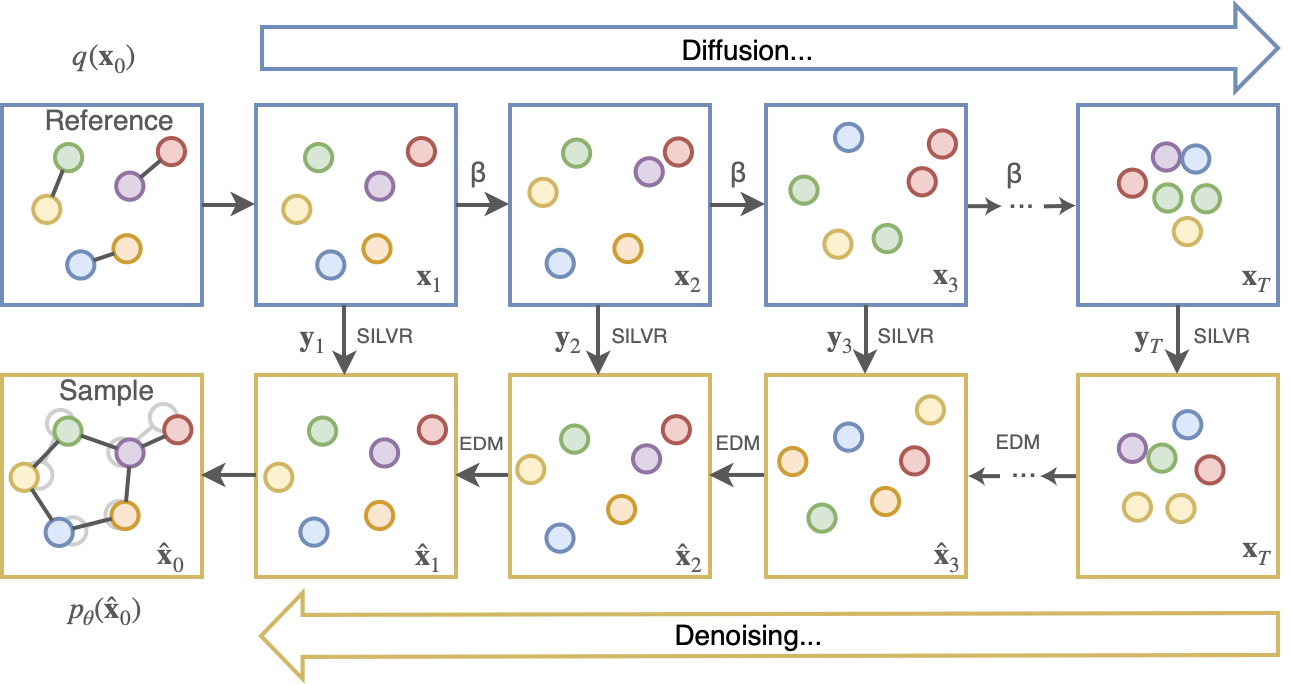}
    \caption{Schematic of the equivariant diffusion model with  selective iterative latent variable refinement (SILVR) indicated for every denoising step. Here the reference in blue on the left is 3 small fragments. They are evolved over time $t$ in the diffusion process to resemble a Gaussian distribution at $t=T$, see Equation~\ref{eqn:noising}. The $\beta$ represents the noise added at each step, and the dots show the omitted steps from time $t=3$ to $t=T$. As atoms effectively 'diffuse' they can be perceived as changing position. To generate a new sample a sample is generated from $p_{\theta}(x)$ according to Equation~\ref{eqn:denoising}, this distribution is from the learned EDM. At each denoising step, a set of reference fragments ($\mathbf{y}_t$) at that same level of noise $t$ is used which is indicated by the SILVR arrows to condition the EDM. This is controlled through SILVR at a given rate $r_{\mathbf{S}}$, until a new sample that resembles the reference is generated (following the bottom row along the yellow boxes and EDM arrows).}
    \label{fig:schematic}
\end{figure}

Denoising probabilistic diffusion models can be separated into two parts, the diffusion process and the denoising process, as shown in the schematic in Figure~\ref{fig:schematic}. These ideas originated in image generation machine learning problems, but can also be applied to molecular generation. A neural network is trained to learn the second part of the process, denoising a Gaussian distribution until an image---or in this case, a molecule---is generated. The idea we propose is to introduce information from a reference molecule at a given rate to the denoising process of a pre-trained model. This is similar to the concept of inpainting~\cite{qin2021image, xie2012image, squires2023artefact}, or more precisely re-painting~\cite{lugmayr2022repaint}, which guides the denoising process at each step towards the reference. In this paper, we show that we can generate new compounds that are similar to a given reference, and link fragments together, using multiple superimposed fragments as input. This method is generalizable to any protein-ligand systems with known fragment hits. For illustration purposes we use the original 23 X-ray fragment hits for the  SARS-CoV-2 Main protease from the COVID Moonshot dataset~\cite{consortium2023open}. In the following, we give an overview of how equivariant diffusion models work and introduce how our method SILVR fits into the framework of an existing pre-trained EDM. We then show, how the SILVR rate, $r_{\mathrm{S}}$ as a modelling parameter can modulate how much of the reference is incorporated in the sample generation and how the original EDM is recovered at  $r_{\mathrm{S}}=0$. As part of this, we illustrate how fragments can be linked using dummy atoms and how newly generated molecules can fit into the existing binding site according to shape complementarity.

\section{Theory}
\label{sec:theory}
\subsection*{Denoising diffusion probabilistic models (DDPM) as generative models}
DDPMs are often used as generative models that were developed for the generation of new images~\cite{sohl-dickstein2015deep, ho2020denoising, nichol2021improved}. More recently, the same idea has also been applied to molecular generation~\cite{hoogeboom2022equivariant}. The main idea behind diffusion models is for a neural network to learn the reverse of a diffusion process, often referred to as denoising, to \textit{sample} a new image or in our case a new molecule. In practice, this is done by training a neural network $\phi$ and generating samples $p(\mathbf{x}_{t-1}|\mathbf{x}_t)=\mathcal{N}(x_{t-1}; \mu_{\theta}(x_t;t), \sigma_t^2\mathbf{I})$ from  a Gaussian transition kernel with learned mean ($\mu_{\theta}(x_t;t)$) and variance ($\sigma_t$), with $\mathbf{x}_t$ being the data noised up to time $t$. Figure~\ref{fig:schematic} shows a schematic of the two main parts of such a DDPM, the diffusion part, where noise is added for each timestep (shown in blue) and the denoising part (shown in yellow) that allows the generation of a new image (molecule).

The \textit{forward} part of the diffusion model is a Markov process: noise is added to a set of data $\mathbf{x}_0$ and over a time interval $t = [1,\ldots,T]$ according to the following distribution:
\begin{equation}
  q(\mathbf{x}_{1:T}|\mathbf{x}_0) = q(x_0)\prod_{t=1}^T q(\mathbf{x}_t|\mathbf{x}_{t-1}).
  \label{eqn:noising}
\end{equation}
The product of conditional probabilities $q(\mathbf{x}_t | \mathbf{x}_{t-1})$ can be modelled as a Gaussian Kernel given by:
\begin{equation}
q(\mathbf{x}_t | \mathbf{x}_{t-1}) = \mathcal{N}(\mathbf{x}_t;\sqrt{1-\beta_t}\mathbf{x}_{t-1}, \beta_t\mathbf{I}),
\end{equation}
where the mean of the normal distribution is given by $ \sqrt{1-\beta_t}$ and $\beta_t$ a fixed variance. 
In order to diffuse directly to timestep $s$ in the diffusion process the following shorthand is possible:
\begin{equation}
    q(\mathbf{x}_t, \mathbf{x}_s)=\prod_{t=1}^s q(\mathbf{x}_t|\mathbf{x}_{t-1})=\mathcal{N}(\mathbf{x}_t|\frac{\alpha_t}{\alpha_s}, \sigma_t^2-\frac{\alpha_t}{\alpha_s}\sigma^2_s),
\end{equation}
for any $t>s$. The parameters $\alpha_t \in \mathbb{R}^{+}$ specifies the amount of retained signal and $\sigma_t \in \mathbb{R}^{+}$ represents the variance and thus the amount of white noise added. The parameter $\alpha_t$ also directly relates to $\beta$ in Figure~\ref{fig:schematic} with $\alpha_t:=1-\beta_t$ and $\langle\alpha\rangle_t:=\prod_{s=1}^t\alpha_s$. $\beta_t$ is a fixed variance schedule which adds noise with each timestep $t$.
Different researchers have examined different noise schedules~\cite{sohl-dickstein2015deep, ho2020denoising}.
What we are actually interested in is learning the reverse process, i.e. the denoising and generating a new sample $\mathbf{\hat{x}}_0$, however the reverse of the process $q(\mathbf{x}_{t-1}|\mathbf{x}_{t=T})$ is intractable. The DDPM learns this reverse transitions $p_{\theta}(\hat{\mathbf{x}}_{t-1}|\mathbf{x}_{t=T})$, which is also a Gaussian transition kernel. This generative (or denoising) process is given by:

\begin{equation}
p_{\theta}(\hat{\mathbf{x}}_{t-1}|\mathbf{x}_{t=T})=\mathcal{N}(\hat{\mathbf{x}}_{t-1}; \mu_{\theta}(\mathbf{x}_t;T), \sigma_t^2\mathbf{I}).
\label{eqn:denoising}
\end{equation}
$\mu_{\theta}$, is the learned mean and $\sigma$ represents a fixed variance for this transition process.
A sample for denoising timestep $t-1$ can be generated from the following equation given the neural network $\phi(\mathbf{x}_t,t)$ that has been trained on the diffusion process: 
\begin{equation}
 \hat{\mathbf{x}}_{t-1} = \frac{1}{\alpha_t}(\mathbf{x}_t-\frac{1-\alpha_t}{\sqrt{1-\langle\alpha_t\rangle}}\phi(\mathbf{\hat{x}}_t,t)) +\sigma \rho,   
\end{equation}

with $\rho\sim\mathcal{N}(0,\mathbf{I})$.

The process is iterated until $t=1$ and as such a new sample $\mathbf{\hat{x}}_0$ of the denoising process---which is intended to represent a proposed molecule design---is generated. 

\subsection*{Equivariant diffusion model (EDM)}
In the previous section we introduced the neural network $\phi(\mathbf{x}_t,t)$. In practice, it makes sense to use an equivariant graph neural network, as it is a data-efficient way to learn about molecules. If a model has rotational and translational equivariance, it means a neural network does not need to learn orientations and translations of molecules from scratch. We chose the EDM by Hoogeboom \textit{et al.}~\cite{hoogeboom2022equivariant} as our baseline generative model, as it provides a generative model for new molecules and has equivariance already built into it. Furthermore, it has all code and pre-trained weights available online at:~\url{https://github.com/ehoogeboom/e3_diffusion_for_molecules}. The basic concept behind equivariance is that the model is invariant to rotations and translation, in this case the E(3) group. This means that scalar (features such as atom types) and vector node properties (such as the positions) will be invariant to group transformations. As a result the order in which a rotation is applied does not matter. The input to the model can be rotated and diffusion/denoising  applied to get a structure, or the diffusion/denoising process can be applied first followed by the same rotation to get the same output. 
Mathematically this means that if we have a set of point $\mathbf{x} = (\mathbf{x}_1,\ldots,\mathbf{x}_N) \in \mathbb{R}^{N\times 3}$ and each of these points has an associated set of scalar feature vectors $\mathbf{h}\in \mathbb{R}^{N\times k}$ which are invariant to group transformations, the position translations and rotations are defined according to the orthogonal matrix: $\mathbf{Rx + t} = (\mathbf{R_{x_1}+t},\ldots, \mathbf{R_{x_N}+t})$. Satorras \textit{et al.}~\cite{satorras2022equivariant} have proposed an E(n) Equivariant Graph Neural Network on which the EDM by Hoogeboom \textit{et al.} builds. 
The actual diffusion process of the Hoogeboom model relies on a set of points $M$ $\{\{\mathbf{x}_i,\mathbf{h}_i\}\}_{i\ldots M}$ for a latent variable that combines atom coordinates $\mathbf{x}_i$ and node features of each atom $\mathbf{h}_i$, such that: $\mathbf{z}_t = [\mathbf{z}_t^{(x)}, \mathbf{z}_t^{(h)}]$. The node features in practice are an array of values containing information such as atom type. These features are encoded using a one-hot encoding. For more details on this see~\cite{hoogeboom2022equivariant}. Based on this latent variable $\mathbf{z}_t$, the diffusion process can be defined similarly to that of equation~\ref{eqn:noising} as:
\begin{equation}
    q(\mathbf{z}_t| \mathbf{x},\mathbf{h}) = \mathcal{N}_{xh}(\mathbf{z}_t|\alpha_t[\mathbf{x,h}],\sigma^2_t\mathbf{I}).
\end{equation}
In the same way, for the generative denoising process the distribution can be written as:
\begin{equation}
    p_\theta(\mathbf{z}_s| \mathbf{z}_t) = \mathcal{N}_{xh}(\mathbf{z}_s | \mu_{\theta t\rightarrow s}( [ \mathbf{\hat{x},\hat{h}}],\mathbf{z}_t),\sigma_{t\rightarrow s}^2\mathbf{I}).
\end{equation}
This is the equivalent of equation~\ref{eqn:denoising}, using $\mathbf{\hat{x}}, \mathbf{\hat{h}}$ as the data variables that are estimated by the neural network. The neural network $\phi(\mathbf{z}_t,t)$ outputs an auxiliary variable $\hat\epsilon = [\hat{\epsilon}^{(x)},\hat{\epsilon}^{(h}]$, from which $\mathbf{\hat{x}}, \mathbf{\hat{h}}$ can be recovered as:
\begin{equation}
    [\mathbf{\hat{x}}, \mathbf{\hat{h}}] = \frac{\mathbf{z}_t}{\alpha_t}-\hat{\epsilon}_t\frac{\sigma_t}{\alpha_t}.
\end{equation}
We use the notation $\kappa$ to generate a sample $\kappa = [\mathbf{\hat{x}}, \mathbf{\hat{h}}]$ from the EDM. 
For more details on the architecture and Hoogeboom's code see~\cite{hoogeboom2022equivariant}. 

\subsection{Iterative Latent Variable Refinement as a conditioning tool}
Conditioning diffusion models is often desirable to, for example, generate similar images to that of original input images, or in the example of Hoogeboom, generate molecules in the presence of an external electric field resulting in more polarizable molecules. However, this conditioning requires retraining the network to accommodate for this condition. Choi \textit{et al.}~\cite{choi2021ilvr} introduced a way to condition DDPMs without having to retrain the neural network. In the generative process it is possible to introduce a condition $c$ using a conditional distribution $p(x_0|c)$:
\begin{equation}
    p_\theta(x_0|c) = \int p_\theta(x_{0:T}|c)dx_{1:T}
\end{equation}
\begin{equation}
    p_{\theta}[x_{\{0:T\}}|c]=p(x_T)\prod_{t=1}^{T}p_{\theta}(x_{t-1}|x_t,c).
\end{equation}
Their trick is to use a reference image $\mathbf{y}$ and place it in the same downsampled filter $\psi(\mathbf{y})$ as the generated image $\psi(\mathbf{\hat{x}}_0)$. This means the original image and the reference are  in the same latent space, so in each denoising step the proposal distribution is matched with that of the reference $\mathbf{y}_t$ noised  to the appropriate timestep. An unconditional distribution at $t$ is generated first:
\begin{equation}
\mathbf{\hat{x}}_{t-1} \sim p_{\theta}( \mathbf{\hat{x}}|\mathbf{x}_t),
\end{equation}
then this  new sample is 'adjusted' according to:
\begin{equation}
    \mathbf{\hat{x}}^{\prime}_{t-1} = \psi(\mathbf{y}_{t-1}) + (I-\psi)(\mathbf{\hat{x}}_{t-1})
\end{equation}
This iterative latent variable refinement (ILVR) means that the condition can be applied during the denoising steps without additional training. In the case of Choi \textit{et al.}~\cite{choi2021ilvr}, they used downsampled reference images as conditioning in order to generate novel images similar to the reference. 

\subsection{Selective Iterative Latent Variable Refinement (SILVR)}
\label{sec:silvr}
We propose a new method, that combines ideas from ILVR by Choi \textit{et al.}~\cite{choi2021ilvr} proposed in an image generation context and the EDM by Hoogeboom \textit{et al.}~\cite{hoogeboom2022equivariant}. This allows us to generate a selective iterative latent variable refinement (SILVR) procedure in which we introduce information of a reference molecule into the denoising process. We describe this method as "selective" due to the ability to guide individual atoms at independent rates. The reference can be a single molecule or a series of superimposed fragments, and additional unguided dummy atoms can also be defined at the beginning of the denoising process. We consider latent space variables $\mathbf{z}=[\mathbf{z}^{(x)},\mathbf{z}^{(h)}]$ for the standard denoising process and  $\mathbf{\tilde{z}} = [\mathbf{z}^{(y)},\mathbf{z}^{(h_y)}]$, for the set of reference coordinates given by $\mathbf{y}$. The vector $\mathbf{h}_y$ contains all the scalar node properties of the equivariant EDM for the reference. 
Similarly to Choi \textit{et al.}, we update the diffusion process at noise level $t$ in the latent space $\mathbf{z}$, with the reference $\mathbf{\tilde{z}}$ using a factor, or vector if used at variable rates for different atoms $r_{\mathrm{S}}$. We call $r_{\mathrm{S}}$ the SILVR rate and this leads to an overall update or conditioning towards a reference at each step in the generative denoising process according to the SILVR equation~\ref{eqn:SILVR}.
\begin{equation}
\mathbf{z}_{t-1}^{\prime} = \mathbf{z}_{t-1}-\alpha_{t-1}r_{\mathrm{S}}\mathbf{z}_{t-1}+r_{\mathrm{S}}\mathbf{\tilde{z}}_{t-1}
    \label{eqn:SILVR}
\end{equation}
As a result, we propose the following algorithm for the generation of condition samples according to SILVR~\ref{alg:silvr}.
\begin{algorithm}
\caption{SILVR}\label{alg:silvr}
\begin{algorithmic}[1]
\State \textbf{Input:} Reference molecule $\mathbf{y,h}_y$, EDM $\kappa$ 
\State \textbf{Output:} Generated molecule $\mathbf{x,h}$

\State Compute $\mathbf{\tilde{z}}_0$ from $\mathbf{y, h}_h$, such that $[\mathbf{z}^{(y)} , \mathbf{z}^{(h_y)} ] = f (\mathbf{y, h}_y)$ is E(3). 
\State Subtract center of geometry (COG) from $\mathbf{\tilde{z}}_0$
\Comment{Center reference at zero}
\State Sample $z_T \sim \mathcal{N}(\mathbf{0,I})$
\For {$t$ = $T,...,1$}
    \State Sample $\boldsymbol{\epsilon} \sim \mathcal{N}(\mathbf{0,I})$
    \State Subtract COG from $\boldsymbol{\epsilon}^{(x)}$
    \State $\mathbf{\tilde{z}}_{t-1} = \alpha_{t-1} \mathbf{\tilde{z}}_{0} +\sigma_{t-1}\times\boldsymbol{\epsilon}$
    \Comment{Noise reference to t=t-1}
    \State $\mathbf{z}_{t-1} \gets \kappa (\mathbf{z}_t,t)$
    \Comment{Compute denoising step}
    \State Update $\mathbf{z}^{\prime}_{t-1} \gets \mathbf{z}_{t-1} - \alpha_{t-1}r_{\mathrm{S}} \mathbf{z}_{t-1} +r_{\mathrm{S}} \mathbf{\tilde{z}}_{t-1}$ \Comment{SILVR equation}
    \State Subtract COG from updated $\mathbf{z}^{\prime}_{t-1}$

\EndFor
\State Add COG($\mathbf{\tilde{z}}_0$) to $\mathbf{z}^{\prime}_0$
\Comment{Move sample to original position of reference}
\State Sample $\mathbf{x,h} \sim p_{\theta}(\mathbf{x,h}|\mathbf{z}^{\prime}_0)$ \Comment{See equation~\ref{eqn:denoising}}
\end{algorithmic}
\end{algorithm}

\begin{figure}
    \centering
    \includegraphics[width=0.5\textwidth]{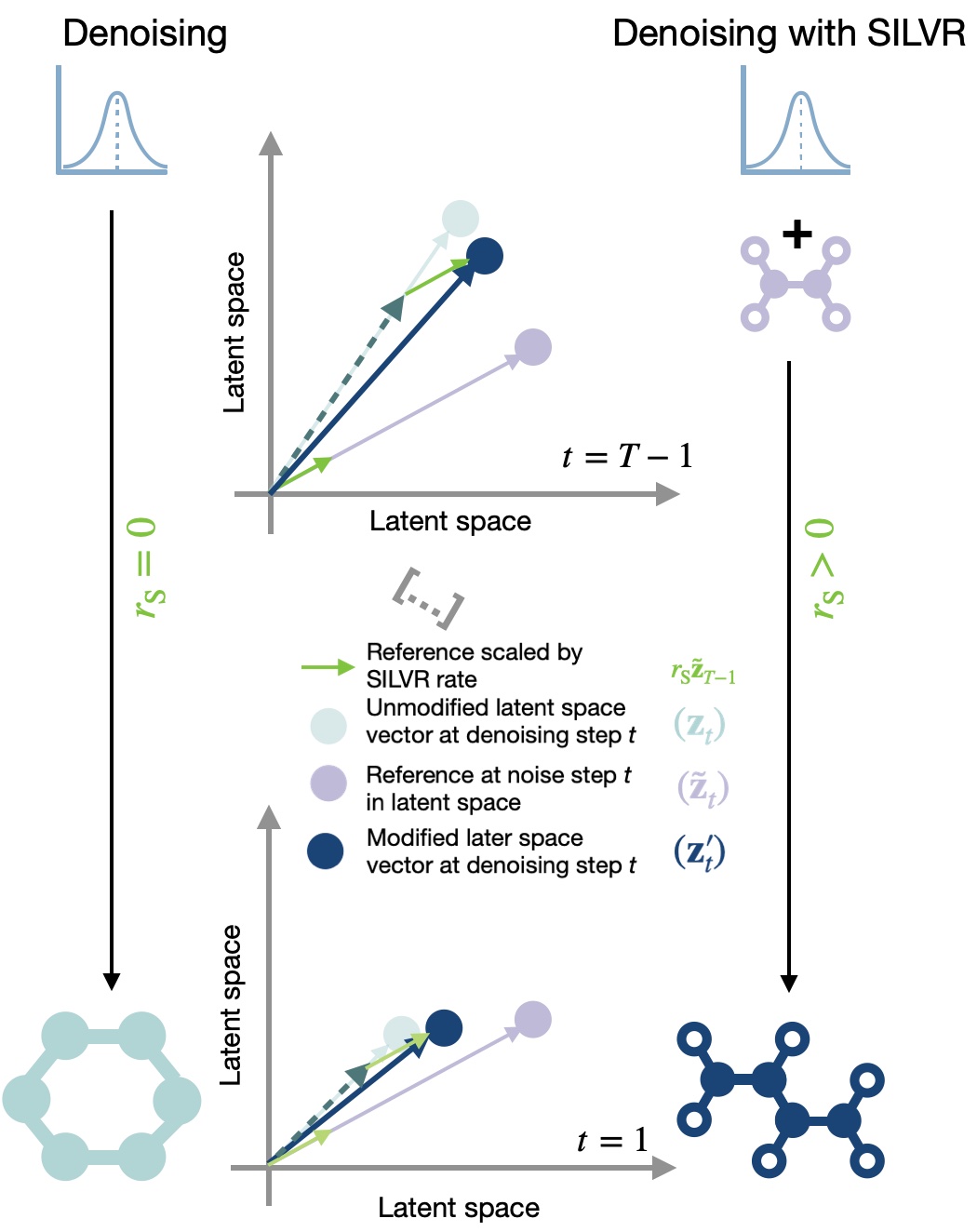}
    \caption{Schematic illustrating the influence of applying the SILVR equation~\ref{eqn:SILVR} to molecules in the latent space. At each denoising step if $r_{\mathrm{S}}>0$ the latent space vector (light blue) gets shifted according to a scaled reference vector (purple) using the SILVR rate (light green). This results in an updated latent vector (dark blue). This is done from the first denoising timestep (top) until the last one (bottom). Repeatedly applying SILVR will result in a molecule that resembles the references for $r_{\mathrm{S}}>0$ (right) and does not for $r_{\mathrm{S}}=0$ (left).}
    \label{fig:silvr_explanation}
\end{figure}

The core of the new method is the addition of a refinement step within the denoising process during runtime of any pre-trained E(3) EDM. The resulting SILVR model produces conditional samples without any conditional training when generating new molecules. Figure~\ref{fig:silvr_explanation} shows an illustrative example of how the SILVR rate $r_{\mathrm{S}}$ is used to shift the latent space vector $\mathbf{z}_{t-1}$ at any point in the denoising process from $T\ldots t=1$, here in 2D for illustration purposes. Using the SILVR equation~\ref{eqn:SILVR}, an existing denoising step (light blue) is brought closer to the reference (purple) in the latent space according to a scaled version of purple using $r_{\mathrm{S}}$ (green).

\section{Methods}
To illustrate the usefulness of this run-time modification, we show how SILVR can be used in the context of fragment-based drug design. The goal is to produce molecules that are complementary to a binding site based on exiting fragments and their set of atomic coordinates. 

\subsection{Conditioning the pre-trained EDM with SILVR}
The EDM by Hoodgeboom \textit{et al.} was trained on the 30 lowest energy conformations of 430,000 molecules from the GEOM dataset with explicit hydrogens~\cite{axelrod2022geom}. One desirable feature of GEOM is that it contains drug-like molecules including 6000 compounds for SARS-CoV-2 targets making this an appealing dataset for generating new molecules that combine or expand fragments for a SARS-CoV-2 target. For more details on the model theory and training refer to Hoogeboom \textit{et al.}~\cite{hoogeboom2022equivariant}. This model strictly only considers atomic coordinates (heavy atoms and hydrogen atoms), while all bond and molecular graph information is ignored. A more recently proposed version of the EDM incorporates molecular graph generation within the denoising diffusion model~\cite{vignac2023midi} improving the quality and potentially also synthetically accessible space of newly proposed molecules. We believe SILVR can be added into the denoising loop of this new model following the same principles we propose here. During training the atomic coordinates, and a one-hot encoding of their element, are passed through a forward diffusion process with iterative addition of Gaussian noise; both the coordinates and one-hot vector get diffused during this process. The extent of noise added at each step is defined by parameter $\beta$ (N.B.\ Figure~\ref{fig:schematic} shows the diffusion process as a Markov chain, however in practice the state at time $t$ can be efficiently computed as a direct transformation of the initial state). The diffusion process is eventually terminated when $t=T$, by which point all structure is lost and all coordinates follow a Gaussian distribution. An equivariant graph neural network (EGNN) is then trained to predict the reverse process, predicting the previous state in the sequence given any state. At runtime, the generative model is instantiated with a sample from a Gaussian distribution and the series of denoising steps are applied resulting in a generated sample consisting of atomic coordinates resembling a drug-like molecule. The resulting Cartesian coordinates can then be interpreted using cheminformatics software to determine the molecular graph. 

The EDM was adapted by introducing SILVR within the denoising process, algorithm~\ref{alg:silvr}, as outlined previously. At runtime, each atom of the reference set of coordinates is mapped to an atom in the EDM; this is achieved by constructing a reference tensor with the same shape as the EDM latent tensor, with the mapping being on a row-by-row basis. The reference coordinates are then translated such that their center of geometry is at the origin and diffused to the same timestep as that of the denoising process. That is, the amount of structure remaining from the reference should match the amount of structure formed by the generative process. A small refinement is applied to add information from the reference coordinates to the latent variable of the denoising process (see line 11 of the algorithm and equation~\ref{eqn:SILVR}). This equation has the effect of shrinking the coordinates towards the origin and then expanding the coordinates out in the direction of the reference, see Figure~\ref{fig:silvr_explanation}. Importantly, the extent of this refinement is defined by the SILVR rate $r_{\mathrm{S}}$, with $r_{\mathrm{S}}=0$ providing no additional refinement and $r_{\mathrm{S}}=1$ resulting in a total replacement of atoms. The diffusion of the reference to $t$ requires the sampling of a Gaussian; at each step of the denoising loop the reference is repeatedly sampled.  Once denoising is complete, sampled molecules can be translated back to the same coordinates as the reference by reverting the initial centre of geometry transformation; in the case of fragment data, this has the effect of returning samples to the binding site of the protein.

By introducing iterative refinement steps, the unconditional EDM can be guided to sample from a smaller region of chemical space that resembles the reference set of coordinates. Figure~\ref{fig:schematic} demonstrates this architecture with the example of three disconnected fragments. Here, the model generates a single 5-membered ring molecule with each atom maintaining the same element, however, notice each atom has drifted slightly from the reference. This is due to the competing effects of SILVR and EDM: the EDM tries to push atoms into a chemically reasonable position, while SILVR pulls atoms towards the reference. The resulting samples, therefore, resemble both valid-looking molecules and the reference set of coordinates. The ability for reference atoms of fragments to move during generation distinguishes SILVR from standard linker design~\cite{guo2023linkinvent, igashov2022equivariant}.

\subsection{Reference Dataset: COVID Moonshot}
Reference molecules were selected from the original 23 non-covalent hits of the SARS-CoV-2 Main Protease (Mpro) identified by the XChem fragment screen~\cite{douangamath2020crystallographic} as part of the COVID Moonshot Project~\cite{consortium2023open, consortium2021open}. A more detailed picture of all fragments is presented in Figure~\ref{fig:moonshot_data} of the Supplementary information (SI). Fragments were visualised using NGLview version 3.03 and combinations were arbitrarily selected as test cases for the following different experimental settings for trying to understand the similarity between the reference and the new sample. We looked at the following scenarios:
\begin{enumerate}
    \item Using three distinct fragments with substantial overlap as a reference to generate a new sample.
    \item Using two fragments with a slight overlap to generate a single new sample.
    \item And using two fragments that are disconnected to investigate linker generation.
\end{enumerate}
Fragments \texttt{x0072} and \texttt{x0354} were selected for benchmarking the effect of the SILVR rate $r_{\mathrm{S}}$ on sampling; \texttt{x1093}, \texttt{x0072}, \texttt{x2193} were randomly chosen to represent three significantly overlapping fragments; \texttt{x0434}, \texttt{x0305} and \texttt{x0072}, \texttt{x0354} were used as partially overlapping fragments; and \texttt{x0874}, \texttt{x0397} were used as two disconnected fragments, resembling a linker design type problem. The bonding information of selected fragments was deleted and Cartesian coordinates were combined into a single \texttt{XYZ} file. Values of $r_{\mathrm{S}}$ were selected and added to the \texttt{XYZ} file to create a reference file containing all experiment setup information. Each experiment was sampled 1000 times. 

\subsection{Different observables were used for the performance assessment of SILVR}

Different observables were used to monitor how realistic and reliable newly generated molecules were and how well they can fit into the existing binding site of Mpro. We looked at the following set of measures:

\subsubsection{Atom stability}
The accuracy of placement of atoms was determined using the stability metric proposed by Hoogeboom \textit{et al.}~\cite{hoogeboom2022equivariant}. This metric infers bonds and bond orders between atoms by considering their interatomic distances. Once all bonds are defined, the valence of each atom is compared to its expected valence and if these values match, the atom is determined as stable. It should be noted that this measure requires the explicit presence of all hydrogens for an atom to be classed as stable. For comparability with other similar published models, this measure was used unmodified. The additional measure of \textit{molecular stability} is often reported together with atom stability (if every atom is stable then the whole molecule is stable), however as has been previously identified, large molecules sampled from this GEOM-trained EDM tend to be unstable.

\subsubsection{RMSD to reference}
The SILVR algorithm creates a one-to-one mapping between reference atom coordinates and sample atom coordinates. The RMSD for this pairwise mapping, ignoring atom identities, was calculated to determine the spatial similarity of samples to the reference. All RMSD calculations were carried out using \url{https://github.com/charnley/rmsd} version 1.5.1.

\subsubsection{Geometric stability - Auto3D}
To determine whether samples represent a true molecular geometry, an independent minimisation of molecular geometries was performed using Atoms In Molecules Neural Network Potential (AIMNet) with Auto3D~\cite{liu2022auto3d}. All samples were read by RDKit and samples containing more than one molecule were removed from the test set. The SMILES string of each molecule was written to a new file, read by Auto3D, and geometry predicted by AIMNet. The RMSD between SILVR-generated coordinates and the Auto3D minimised coordinates were calculated with RDKit version 2022.03.5.

\subsubsection{Shape complementarity of the generated sample to the protein}
The agreement in the shape of samples and the MPro binding site was determined using openeye-toolkit version 2022.2.2 Gaussian scoring function \textit{shapegauss}~\cite{openeyescientificsoftwareinc.oedocking,kelley2015posit}. This scoring function measures the shape complementarity between the ligand and receptor by considering each heavy atom as a Gaussian function~\cite{mcgann2003gaussian}. The most favourable score occurs “when two atoms touch but do not overlap”. This metric does not consider any intermolecular interactions beyond shape complementarity. The protein receptor file was prepared from the Mpro-x0072 crystal structure with removal of the ligand. The \texttt{XYZ} coordinates of samples were directly read into the openeye-toolkit and the pose was re-scored with \textit{shapegauss}.

\section{Results}
In the following, we will demonstrate how SILVR can be used to generate conditioned samples to a reference using a pre-trained EDM without additional training. The main questions we set out to answer with SILVR were:
\begin{enumerate}
    \item Can we generate samples from the EDM that are similar to the reference structures?
    \item Is there a SILVR rate $r_{\mathrm{S}}$ that will provide enough diversity while still retaining reference features?
    \item Do the generated samples of new molecules still fit into the Mpro binding site?
    \item Can we link molecule fragments without incorporating binding site information as part of additional training?
\end{enumerate}

\subsection{The SILVR rate $r_{\mathrm{S}}$ effectively modulates similarity to the reference structures}
Qualitatively, the generated molecular samples from the conditioned EDM  show a clear resemblance to their reference structures, with similarity increasing with $r_{\mathrm{S}}$. Figure~\ref{fig:examples_curated}, in the SI, shows two example samples started from fragments \texttt{x0072} and \texttt{x0354} over a range of SILVR rates between $r_{\mathrm{S}}=0$ to $r_{\mathrm{S}}=0.02$. As expected at no conditioning random samples are generated that do not resemble the reference fragments.  At low values of $r_{\mathrm{S}}$ ($< 0.0025$) the sampled molecules only show an approximate agreement in orientation. At medium values of $r_{\mathrm{S}}$ ($0.0025 \le r_{\mathrm{S}} < 0.01$) the resulting samples begin to produce key structural features such as ring systems and heteroatoms at positions seen in the reference. At high values of $r_{\mathrm{S}}$ ( $\ge 0.01$) the resulting samples have a very high resemblance to the reference with most structural features in correct positions, however, the diversity of samples is significantly reduced and structures start to become chemically less reasonable. At very high values of $r_{\mathrm{S}}$ ($> 0.02$) there is a very high similarity between samples and the reference, however, most structures no longer resemble valid molecules. The best molecules are formed at intermediate values of $r_{\mathrm{S}}$ ($0.05 \le r_{\mathrm{S}} \le 0.01$) offering a trade-off between similarity to the reference, sampling diversity, and molecular likeness. This is further validated by looking at stability measures. 

\begin{figure}[ht!]
    \centering
    \includegraphics[width=0.8\textwidth]{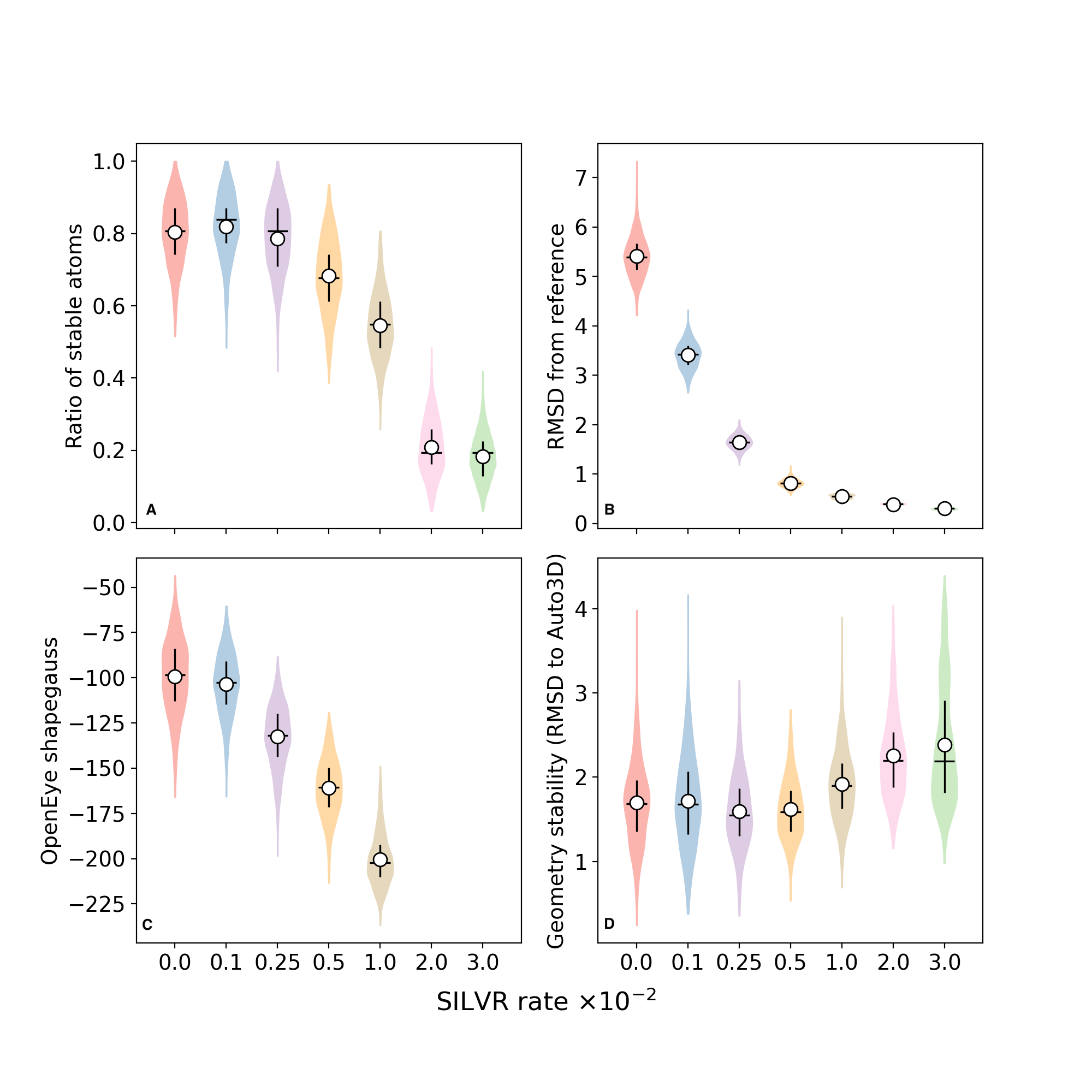}
    \caption{Validation measures of SILVR model using fragments \texttt{x0072} and \texttt{x0354} as reference coordinates. A: Ratio of stable atoms - an atom is determined as stable if the valence matches the expected valence for the element B: RMSD from reference - the calculated RMSD between the reference and sample, using an absolute one-to-one mapping ignore atom identity with low RMSD meaning molecules are similar to the reference and high RMSD they are not. C: OpenEye measure \textit{shapegauss} - a Gaussian scoring function describing the shape fit between Mpro and samples, ignoring chemical interactions. A lower score means a better shape fit of the molecule D: Geometry stability - AIMNet geometry optimisation was completed with Auto3D using the SMILES string of each sample. RMSD was calculated between the predicted geometry and the sampled geometry using RDKit. Horizontal Lines indicated the sample median and circles the sample mean.}
    \label{fig:measures}
\end{figure}
\subsection{Intermediate SILVR rates produce stable and varied molecules}
To assess the stability and variability of generated molecules we looked at four different metrics, as discussed in the methods section. We generated 1000 samples at different $r_{\mathrm{S}}$ using fragments \texttt{X0072} and \texttt{X0354} as a reference. Figure~\ref{fig:measures} summarises the findings from these experiments with violin plots generated across the 1000 samples. The zeroth test we made with the generated samples was looking at how many generated molecules were fragmented i.e. are not a single connected molecular graph, with respect to $r_{\mathrm{S}}$. This is presented in Figure~\ref{fig:fragmentation} in the SI. At an intermediate $r_{\mathrm{S}} = 0.025$ just over 50\% of the generated samples are not fragmented meaning that, only one in two generated molecules can be analysed further. The subsequent analysis is carried out only on whole molecular graphs. 
Figure~\ref{fig:measures} A looks at the \textit{atom stability} measure as introduced by Hoogeboom \textit{et al.}~\cite{hoogeboom2022equivariant}. Samples generated at low $r_{\mathrm{S}}$ tend to have similar atom stability, samples start becoming less stable around $r_{\mathrm{S}}=0.005$, and become totally unstable at $r_{\mathrm{S}}=0.02$. This trend can largely be explained due to issues around hydrogens. The atom stability measure calculates whether the valance of each atom matches what is expected for that atom, however, the measure requires the presence of explicit hydrogens. A carbon skeleton with appropriate C-C bond lengths would be determined as \textit{unstable} unless each carbon was populated with explicit hydrogens. In the case of high $r_{\mathrm{S}}$ values, the SILVR method pulls atom types strongly towards the reference. Since there are no hydrogens in the reference, all atoms are mapped to heavy atoms, and therefore most atoms are unable to satisfy a full valence. Adding hydrogens explicitly to the molecules, through OpenBabel or RDkit, is a way of improving this measure.

The similarity of samples to their reference set of coordinates was determined by RMSD, with a clear inverse correlation observed between $r_{\mathrm{S}}$ and RMSD, as seen in Figure~\ref{fig:measures} B. This indicates that the extent of guidance of atoms towards the reference set of coordinates can be fine-tuned by varying $r_{\mathrm{S}}$. 

The next test we carried out was to determine whether the sampled molecular geometries are reasonable. For this purpose a separate geometry optimisation protocol was devised using the SMILES strings of the generated molecules and the RMSD between the generated molecule and the geometrically optimised molecule calculated. The results of this are found in Figure~\ref{fig:measures} D. At low to medium values of $r_{\mathrm{S}}$ ($< 0.01$), the average RMSD values all fall in the $1.6 - 1.7$ \AA~  range. Importantly, no difference is seen between the control set ($r_{\mathrm{S}}=0$) and the SILVR samples ($0 < r_{\mathrm{S}} < 0.01$) indicating the quality of generated molecules was not impeded by the SILVR protocol. 
 The synthetic accessibility (RDKit SA score~\cite{ertl2009estimation}) (Figure~\ref{fig:SA_QED} A) and the Quantitative Estimation of Druglikeness (QED) (Figure~\ref{fig:SA_QED} B) were also estimated as shown in Figure~\ref{fig:SA_QED} in the SI. SILVR does not affect the SA score, or massively change QED for $r_{\mathrm{S}}$ with the best outcomes. 
 
\subsection{Generated samples with SILVR fit the binding site of Mpro}
As one of the main motivators for SILVR is to be able to generate new molecules that fit directly into a binding site based on input fragments, we measured shape complementarity between newly generated samples and the Mpro binding site. We used the OpenEye \textit{shapegauss} scoring function for this purpose~\cite{kelley2015posit}. From Figure~\ref{fig:measures} C it can be seen that the shape complementarity of samples improves with increasing $r_{\mathrm{S}}$, demonstrating that SILVR can produce ligands of binding site geometry when guided by the coordinates of fragment molecules. The lack of data for $r_{\mathrm{S}}=0.02$ and $r_{\mathrm{S}}=0.03$ was because all \textit{shapegauss} calculations failed. We believe this was due to the atom coordinates representing highly strained and internally clashing molecules, and the scoring algorithm either failed to read the molecules or identified them as bad conformations.

\begin{figure}[!ht]
    \centering
    \includegraphics[width=0.8\textwidth]{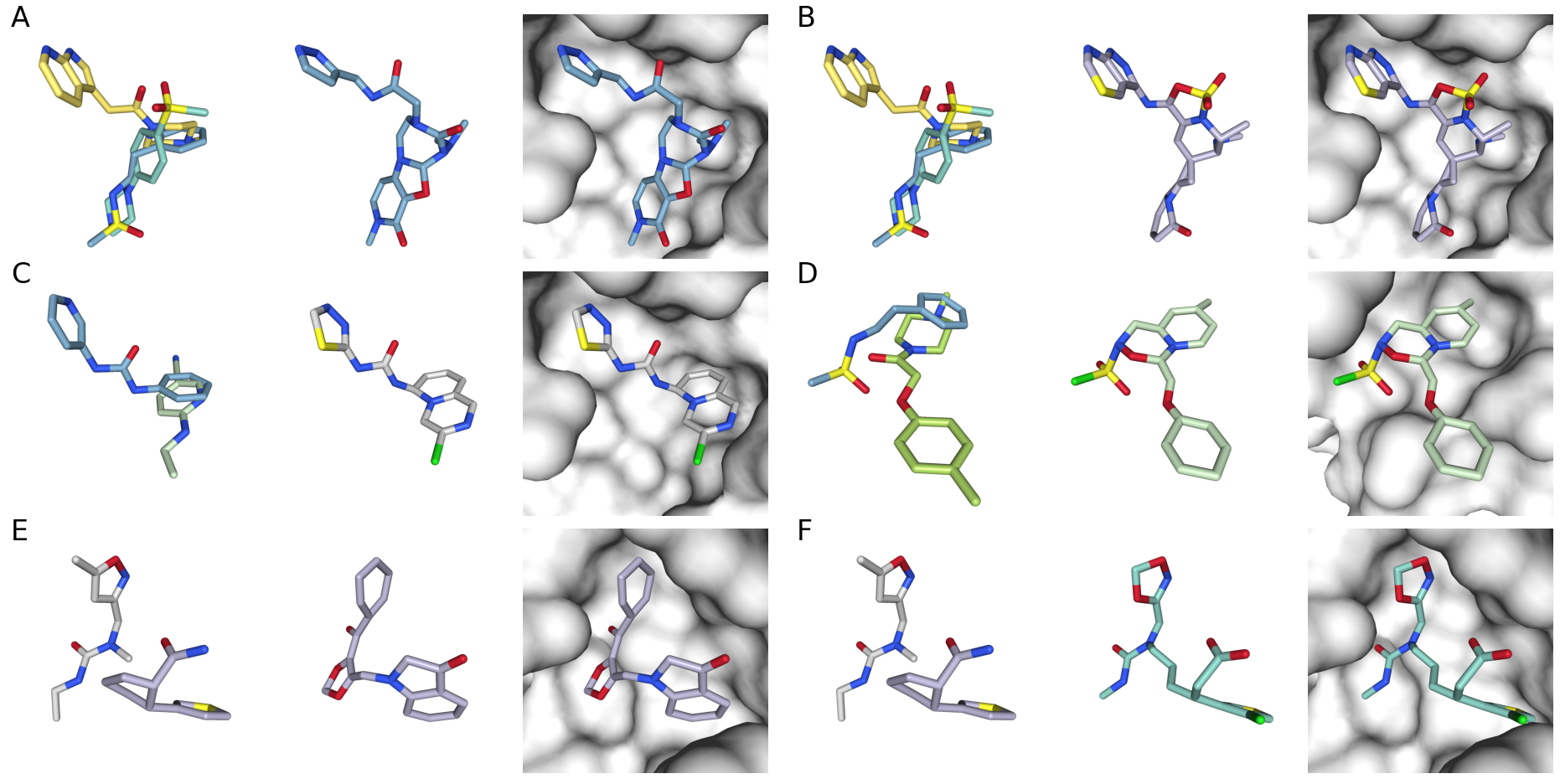}
    \caption{Examples of generated molecules from different experiments testing different overlap models. The reference fragments used as input to SILVR  are shown in the left column, the sampled molecule in the middle column, and the sampled molecule translated to the protein binding site in the right column. The left set of samples (A,C,E) has a SILVR rate $r_{\mathrm{S}}=0.005$, and the right set of samples (B, D, F) $r_{\mathrm{S}}=0.01$. Row 1: Three significantly overlapping fragments A and B (\texttt{x1093, x0072, x2193}). Row 2: Two fragments partially overlapping C (\texttt{x0434, x0305}), D (\texttt{x0072, x0354}). Row 3: Two disconnected fragments E and F (\texttt{x0874, x0397}) samples generated including 10 dummy atoms (method described in SI). The selection of molecules was hand-curated.}
    \label{fig:samples}
\end{figure}

In addition to the experiments using two fragments and generating 1000 samples, we also looked at different combinations of fragments and resulting molecules. In general, the trends of Figure~\ref{fig:measures} were preserved for all experiments. In the following, we present three cases we investigated in detail, the test case using 3 fragments as a reference with substantial overlap, using two fragments with some overlap, and two disconnected fragments for the linker generation experiment. 

Figure~\ref{fig:samples} A-F shows hand-curated samples for the 3 different scenarios for $r_{\mathrm{S}}=0.005$ (A,C,E) and $r_{\mathrm{S}}=0.01$ (B,D,F).

Figures~\ref{fig:samples} A and~\ref{fig:samples} B demonstrate a superposition of three significantly overlapping fragments that result in generated molecules that fit the Mpro binding site well. Scrutinising sample A with $r_{\mathrm{S}}=0.005$, we can see the azaindole fused ring system has been interpreted as a pyrrole ring, the ketone transformed into an amide (maintaining the same carbonyl position), the sulfonyl group vanished, and the overlapping atoms have transformed into a fused ring system. As a whole, the general geometry of the sample reflects the reference, however, functional groups are only weakly preserved. In contrast, sample B presents the same reference set but with $r_{\mathrm{S}}=0.01$. This new sample maintains the same geometry but better preserves key functional groups: the fused ring system is the same size, and satisfyingly the carbonyl oxygen has merged with the sulfonyl group to form a cyclic sulfamate ester.

Figure~\ref{fig:samples} C shows a merger of two partially overlapping fragments with $r_{\mathrm{S}}=0.005$. While the urea group was successfully preserved, the 6-membered ring shrank to a 5-membered heterocycle. Of particular interest is the formation of the fused ring system. At first glance, it might be assumed that reference atoms map to the sample atom closest in space, however in actuality they travel up to 1.7 \AA~to arrive at their final position (Figure~\ref{fig:atom_displacement} in the SI) In this case, the nitrogen atoms observed in the fused ring are directly obtained from the nitrogen atoms in the reference, however, their final position is one bond’s length from their reference. This shows the flexibility of each sample atom to explore within a radius (defined by $r_{\mathrm{S}}$) of the reference atom. The fact that the sample molecule populates a similar region of space to the reference is the result of the aggregate effects of each mapping, as opposed to the strict fixation of each atom. 

In contrast, Figure~\ref{fig:samples} D shows a stricter merging of two fragments, with $r_{\mathrm{S}}=0.01$. Visibly, the scaffold of the lower fragment has been maintained while the top fragment has contributed to a fused ring. Interestingly, the sulfonamide and carbonyl (from opposing fragments) have merged to form N-oxazinane sulfonyl chloride, demonstrating a particularly creative result from SILVR. 

\subsection{Fragments can be linked using SILVR and additional dummy atoms}
\label{sec:linker}
Being able to reliably link fragments that sit in a binding site of a protein is crucial for the design of potential new drugs. Here we demonstrate how this can be done  without retraining and no training for the specific task of linker design~\cite{huang20223dlinker,imrie2020deep}. Using conditioning through SILVR allows the generation of linkers between fragments without the need to retrain the EDM. This is illustrated in the example of Figure~\ref{fig:samples}E and F. While it was possible to use SILVR as described in the theory section, better results were obtained with the addition of \textit{dummy atoms}. These are atoms which are present in the EDM without a mapping to a reference atom, and so are free to explore the whole coordinate space without guidance from SILVR. The successful implementation of dummy atoms requires a slight modification of the SILVR algorithm and is outlined in the SI. The results of these experiments continue the same trends previously observed where the $r_{\mathrm{S}}=0.005$ produces samples of approximate geometric similarity, whereas $r_{\mathrm{S}}=0.01$ produces a more strict mapping, with a clearly preserved urea group, a slightly modified ring system, and an amide interpreted as a carboxylate. When varying the number of dummy atoms used for linker generation, the atoms stability measure is not impacted for $r_{\mathrm{S}}=0.005$, as seen in Figure~\ref{fig:dummy_atoms} A and for $r_{\mathrm{S}}=0.01$ (Figure~\ref{fig:dummy_atoms} B) it only marginally improves with more dummy atoms found in the SI. Using a better EDM which resolves explicit and implicit hydrogens better will likely improve this more. 


\section{Discussion and outlook}
\label{sec:discussion}
SILVR, as presented, represents a method in which a general equivariant diffusion model (EDM) can be conditioned to generate samples that resemble a reference structure, without any additional training needed. We showed that SILVR can complete both fragment merging and linking type tasks, without any \textit{a priori} knowledge of these design challenges. Considering all results with respect to the control EDM ($r_{\mathrm{S}}=0.0$), we show that at intermediate values of $r_{\mathrm{S}}$ the SILVR protocol produces molecules of equal quality to that of the unmodified EDM, while also guiding molecules towards reference structures. We, therefore, claim that if a diffusion model can be successfully trained to produce random high-quality drug-like structures, SILVR will provide molecular designs from desired regions of chemical space without harming the quality of molecules. Our method poses a direct interface between crystallographic fragment data and \textit{de novo} molecular generation. There are a few ways in which the current method can be improved further, but we deem these out of scope for this work. 

\subsection{The number of unfragmented molecules generated can be improved}
The samples generated by SILVR are often of poorer quality than the samples selected in Figure~\ref{fig:samples}. Across all samples around half of the samples were determined by RDKit to be fragmented, meaning the sample contained two or more distinct molecular graphs (See the uncurated list of samples in Figure~\ref{fig:uncurated_sample} in the SI). It was observed qualitatively that fragmented samples typically contained corrupted structures (multiple fragmentations, linear carbon chains, flattened rings, etc). We believe this fragmentation is triggered during intermediate steps of denoising, resulting in an unstable latent representation and subsequently poor EDM inference. Fragmentation becomes a particular issue for linker design type SILVR tasks (Figure~\ref{fig:samples} E and F), where the reference coordinates direct the latent variables away from each other, triggering fragmentation. For these experiments, 65\% of all samples were fragmented. Further work is needed both with EDM and with SILVR to reduce these rates of fragmentation. 

\subsection{The synthetic accessibility of the underlying EDM has a direct impact on the generated molecules}
 For our experiments, the synthetic accessibility of SILVR-generated molecules resembles the performance of the unmodified EDM. In order to achieve synthetically accessible samples with SILVR an improved EDM will need to be designed. An improved version of the EDM we have used has recently been proposed using more explicit information on bond order and represents the next appropriate step for testing SILVR~\cite{vignac2023midi}.

\subsection{The retention of functional groups from the reference structure is challenging}
When applied in a drug-design context, the conservation of key functional groups in exact spacial positions is crucial to maintain desirable protein-ligand interactions. The series of molecules in Figure~\ref{fig:examples_curated} of the SI shows a loss of the sulfonyl chloride group present in the reference, which may be undesirable. This issue can be solved by changing $r_{\mathrm{S}}$ from a scalar to a vector ($\mathbf{r}_{\mathrm{S}}$), and by assigning particularly high $\mathbf{r}_{\mathrm{S}}$ values to selected atoms of the reference. Optimisation of $\mathbf{r}_{\mathrm{S}}$ vectors for actual drug design applications may become viable with a more suitably trained EDM.


\subsection{Placements of hydrogens and dummy atoms needs additional trials}
An EDM with explicit hydrogens will improve the overall models. At the moment there is a mixing of explicit and implicit hydrogens depending on the need for analysis and input. An optimal model can account for hydrogens both explicitly and implicitly allowing for scoring of either. In addition using dummy atoms strategically for growing certain parts of a fragment is something to be explored further in the future.

\section{Conclusions}
We developed SILVR, a method that can be injected into a pre-trained equivariant diffusion model that serves as a molecular generator to explore new chemical space. SILVR allows the conditioning of molecules based on a reference set of molecules, e.g. a fragment from an X-ray fragment screen. The SILVR rate $r_{\mathrm{S}}$ allows the tuning of 'how much of the reference' molecule should be taken into account when generating new molecules, with medium values of $r_{\mathrm{S}}$ around 0.005 to 0.01 giving the best results. The simple conditioning against a reference set of molecules means that the model can be used for tasks of fragment linker design, as well as generating new molecules that fit into an existing binding pocket without any specific training needed towards these tasks. This method is also generalizable, as it works on ligands with no information needed from the protein in its current form. In the future, given improvements in EDMs that produce more realistic and synthetically accessible molecules, this method can cheaply generate structures exploring new chemical space with desired conditioning towards existing fragment hits.

\section{Data Availability}
All data for the experiments carried out and instructions on how to reproduce this work can be found at~\url{https://github.com/meyresearch/SILVR}. An updated version of Hoogeboom \textit{et al.} EDM that includes SILVR can be found at~\url{https://github.com/nichrun/e3_diffusion_for_molecules}.


The authors thank Matteo T. Degiacomi and John D. Chodera for useful discussions and feedback on the manuscript.

\printbibliography

\newpage 
\begin{appendix}

\section{Summary of additional figures }
\begin{figure}
    \centering
    \includegraphics[width=0.8\textwidth]{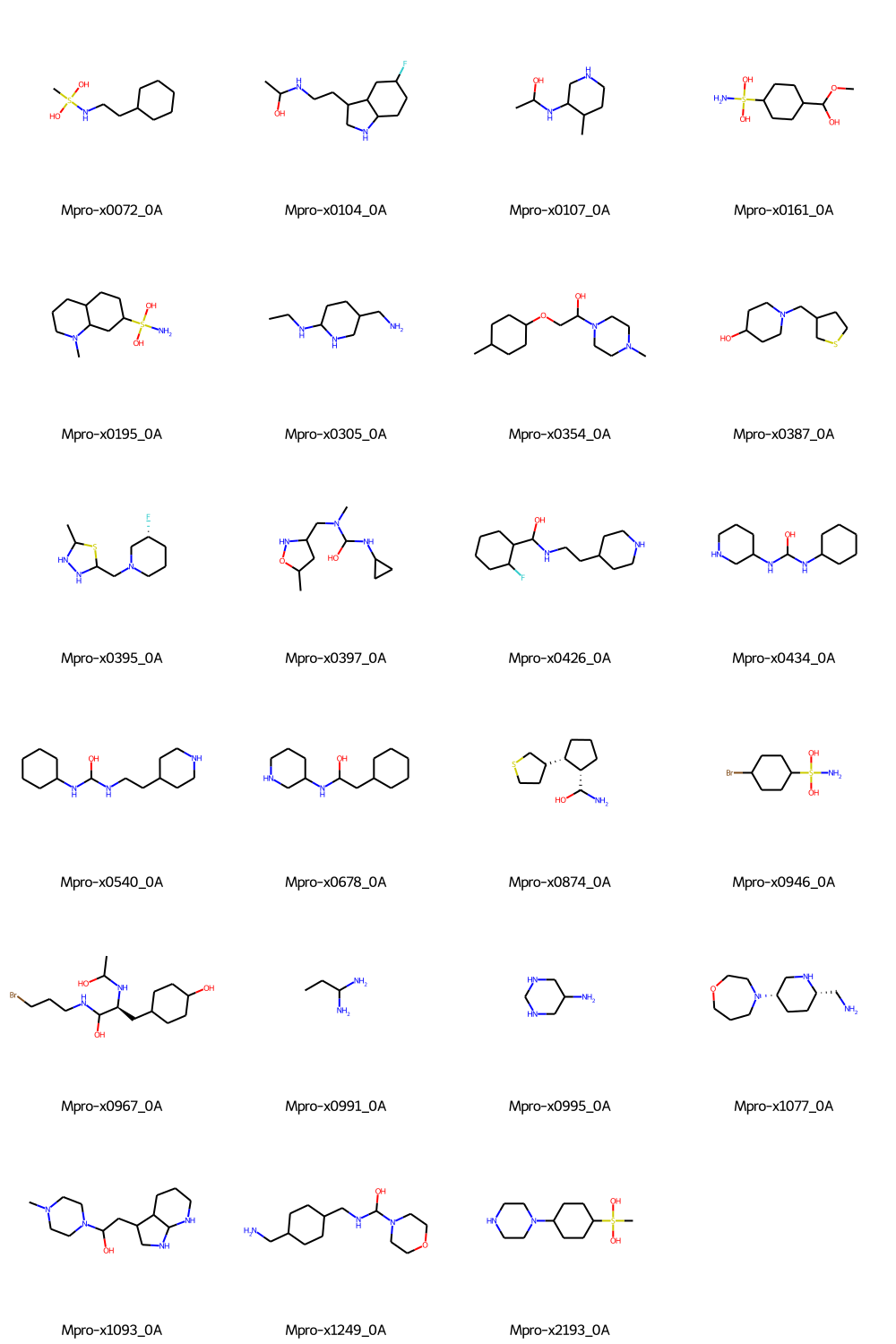}
    \caption{2D Mpro structures from the moonshot dataset.}
    \label{fig:moonshot_data}
\end{figure}

\begin{figure}[hb!]
    \centering
    \includegraphics[width=\textwidth]{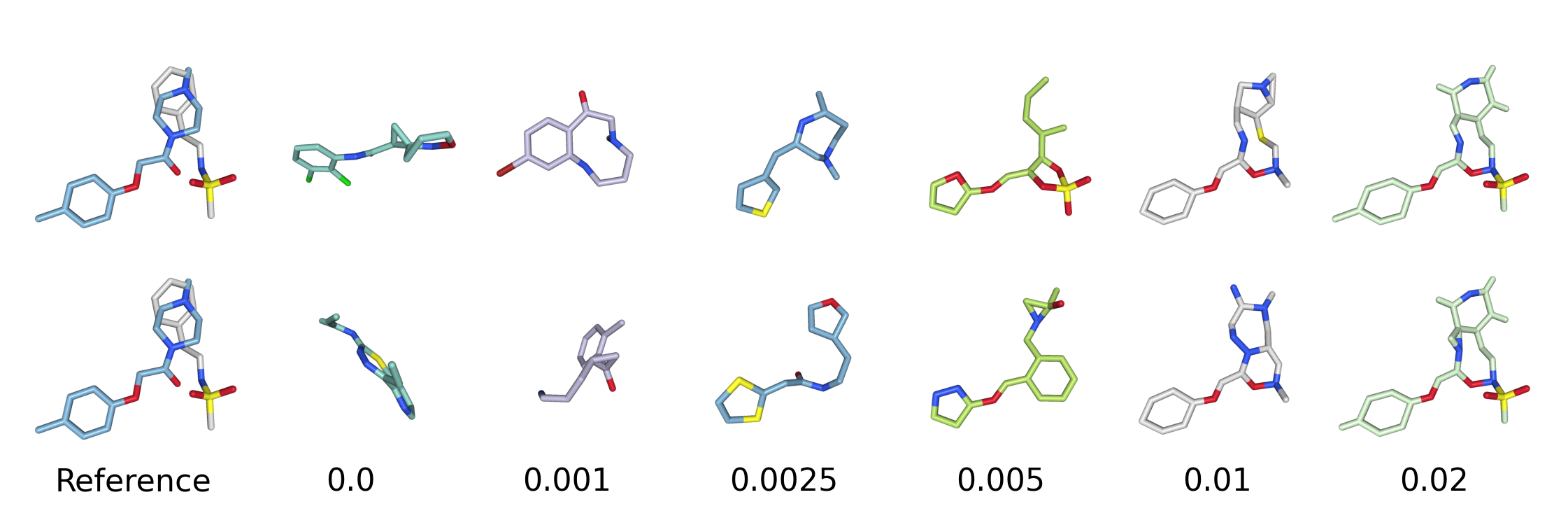}
    \caption{Two random samples from the same reference using different SILVR rates. All bonds were inferred from \texttt{XYZ} coordinates with OpenBabel. All bonds were visualised as single bonds and hydrogen atoms were deleted for clarity. Increasing SILVR rate results in sampled atom coordinates coming closer in space, and element type, to the reference while still resembling a truly molecular structure.}
    \label{fig:examples_curated}
\end{figure}

\begin{figure}
    \centering
    \includegraphics[width=0.5\textwidth]{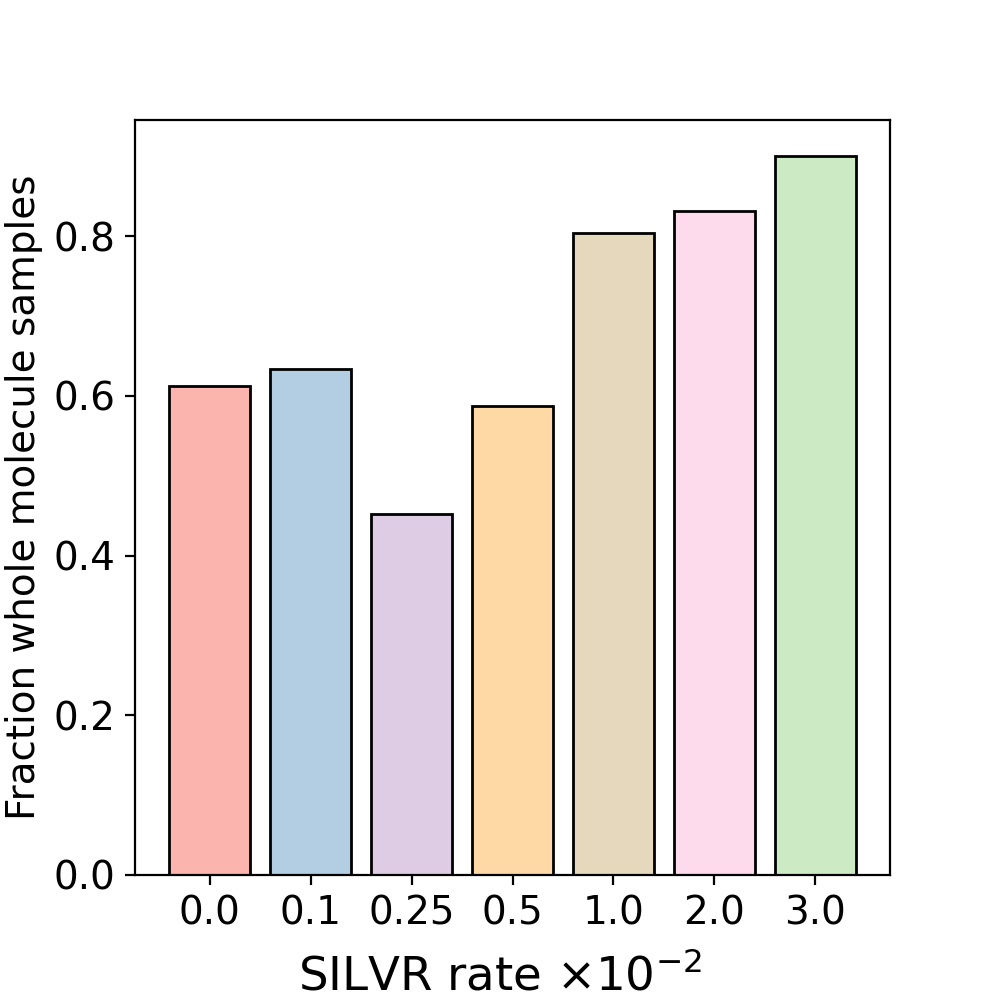}
    \caption{Fraction of molecules not fragmented with respect to the SILVR rate}
    \label{fig:fragmentation}
\end{figure}

\begin{figure}
    \centering
    \includegraphics[width=0.8\textwidth]{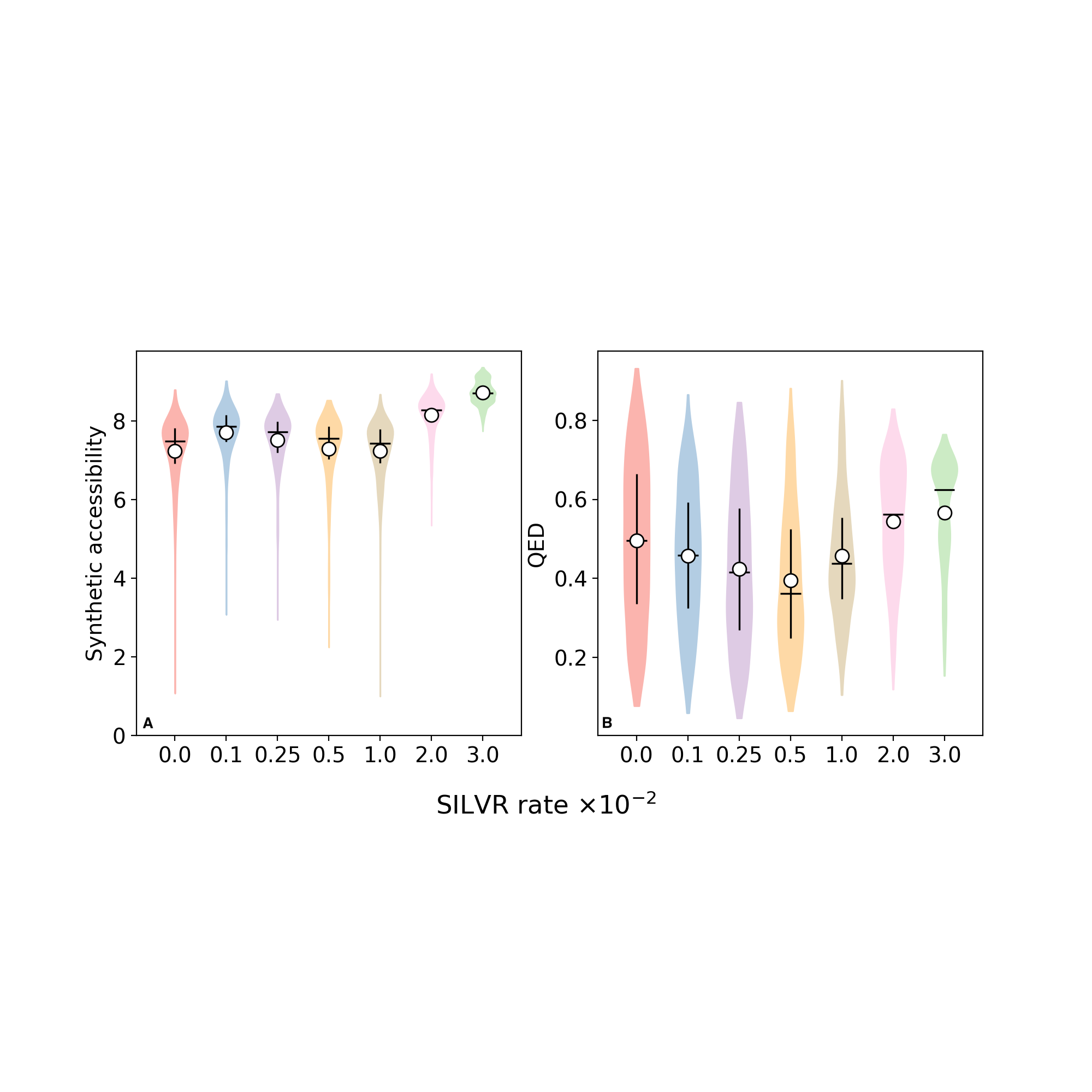}
    \caption{A: The synthetic accessibility score (SA score) estimates the synthetic feasibility of molecules based on fragments contributions.~\cite{ertl2009estimation} This was calculated for all non-fragmented samples using an RDKit implementation of the scoring function SAScorer~\cite{ertl2009estimation}. A lower score indicates an easier-to-synthesise molecule: most catalogue and bioactive molecules fall in the range 2-5, while a score greater than 7 represents the upper end of complexity for natural products.~\cite{ertl2009estimation} B: The Quantitative Estimate of Druglikeness (QED) score combines a selection of descriptors such as molecular weight and calculated Log(P) to estimate the drug-likeness of a molecule.~\cite{bickerton2012quantifying, wildman1999prediction} This was calculated using RDKit with default settings across all non-fragmented samples. A higher score indicates a more drug-like molecule.}
    \label{fig:SA_QED}
\end{figure}

\begin{figure}
    \centering
    \includegraphics[width=0.8\textwidth]{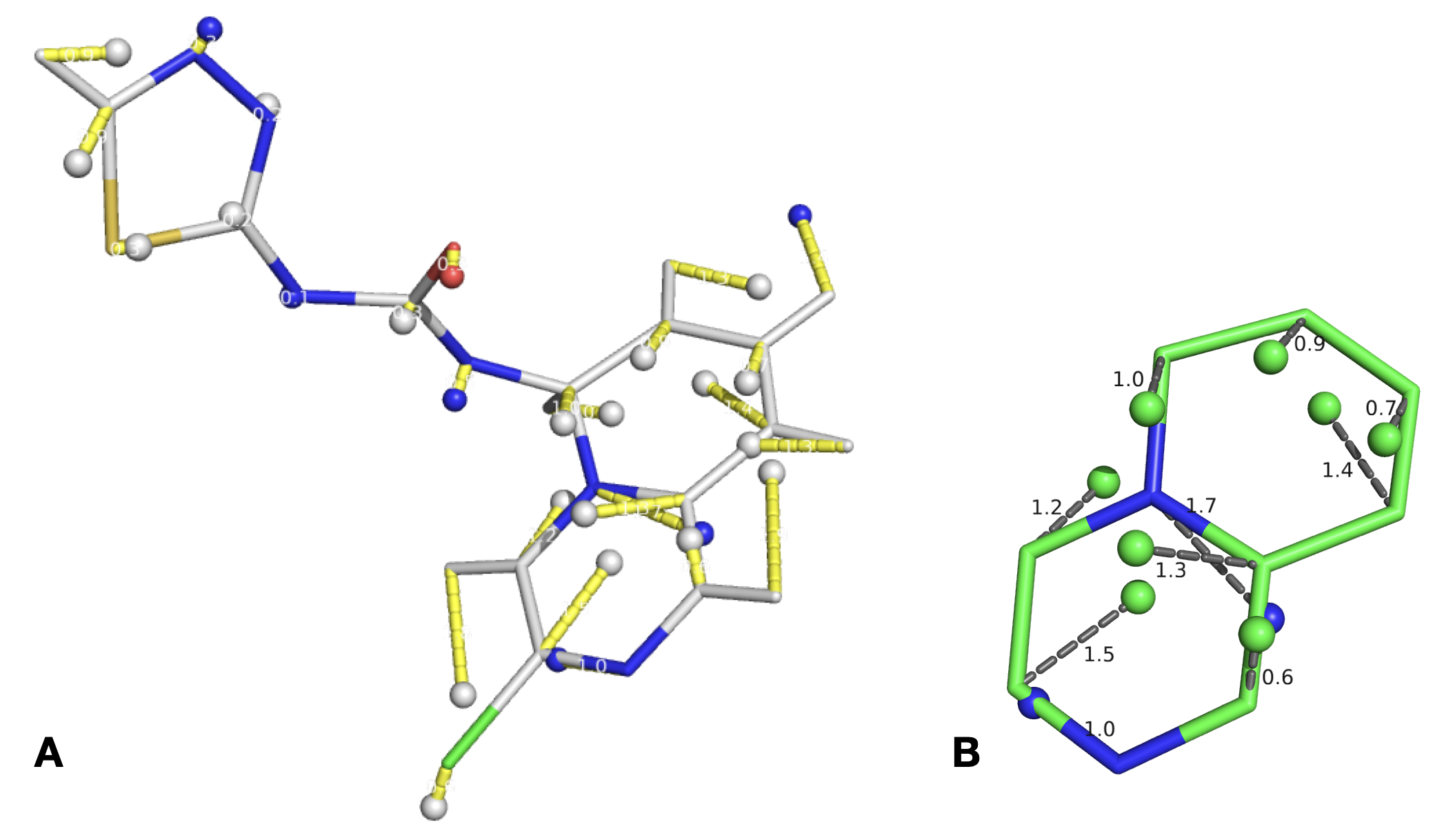}
    \caption{Example looking at the displacement of reference atoms after SILVR denosing. A: for the whole molecule. B: Zoom in, on the displacement of atoms in the 9aH-Pyrido[1,2-a]pyrazine heterocycle. }
    \label{fig:atom_displacement}
\end{figure}

\begin{figure}
    \centering\includegraphics[width=0.8\textwidth]{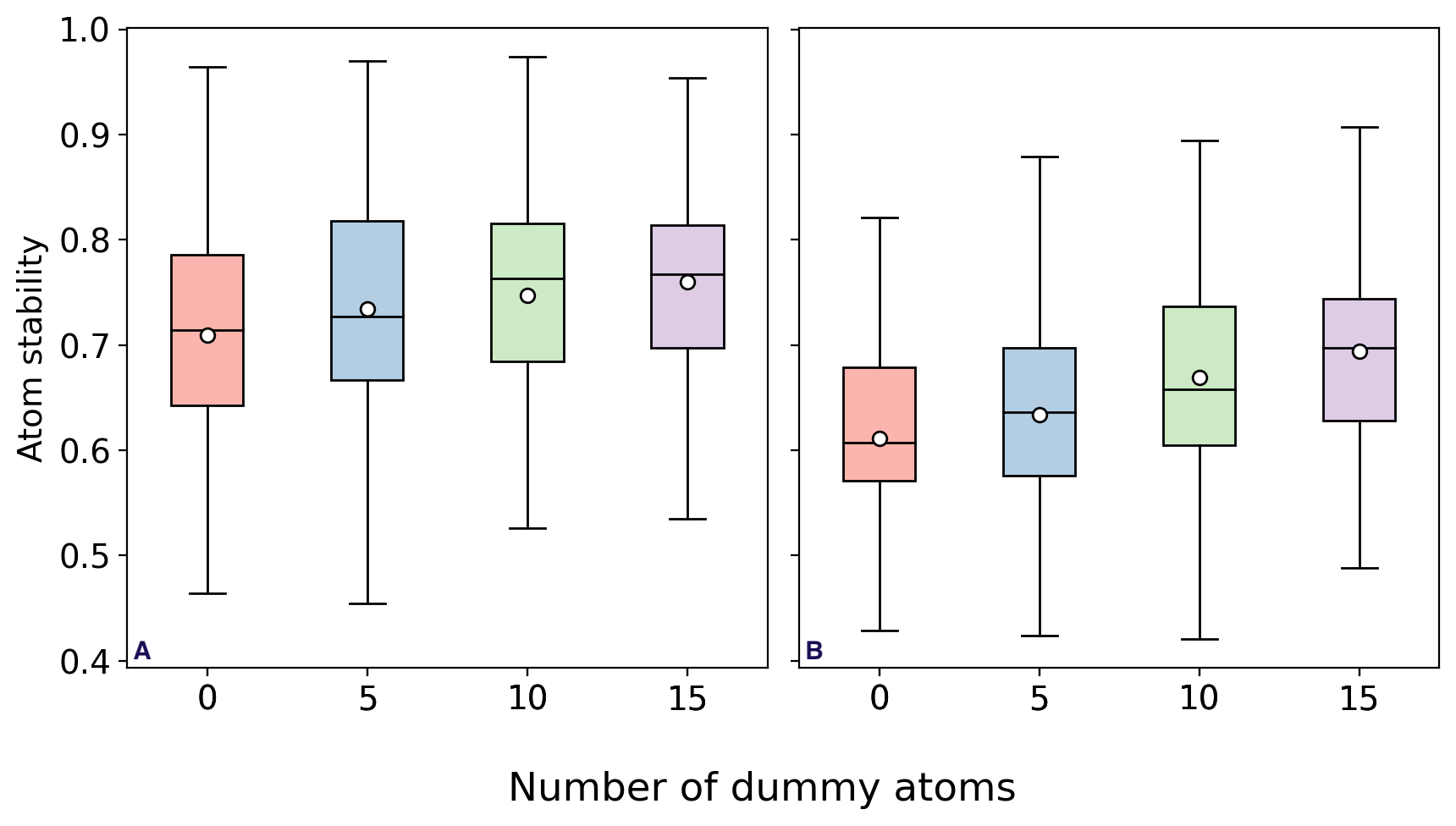}
    \caption{Effect of dummy atoms on the atom stability for linker design type experiment with reference fragments x0874 and x0397. A: $r_{\mathrm{S}} = 0.005$ and B:$r_{\mathrm{S}} = 0.01$. Lines represent sample median, circles sample mean.}
    \label{fig:dummy_atoms}
\end{figure}

\begin{figure}
    \centering
    \includegraphics[width=0.8\textwidth]{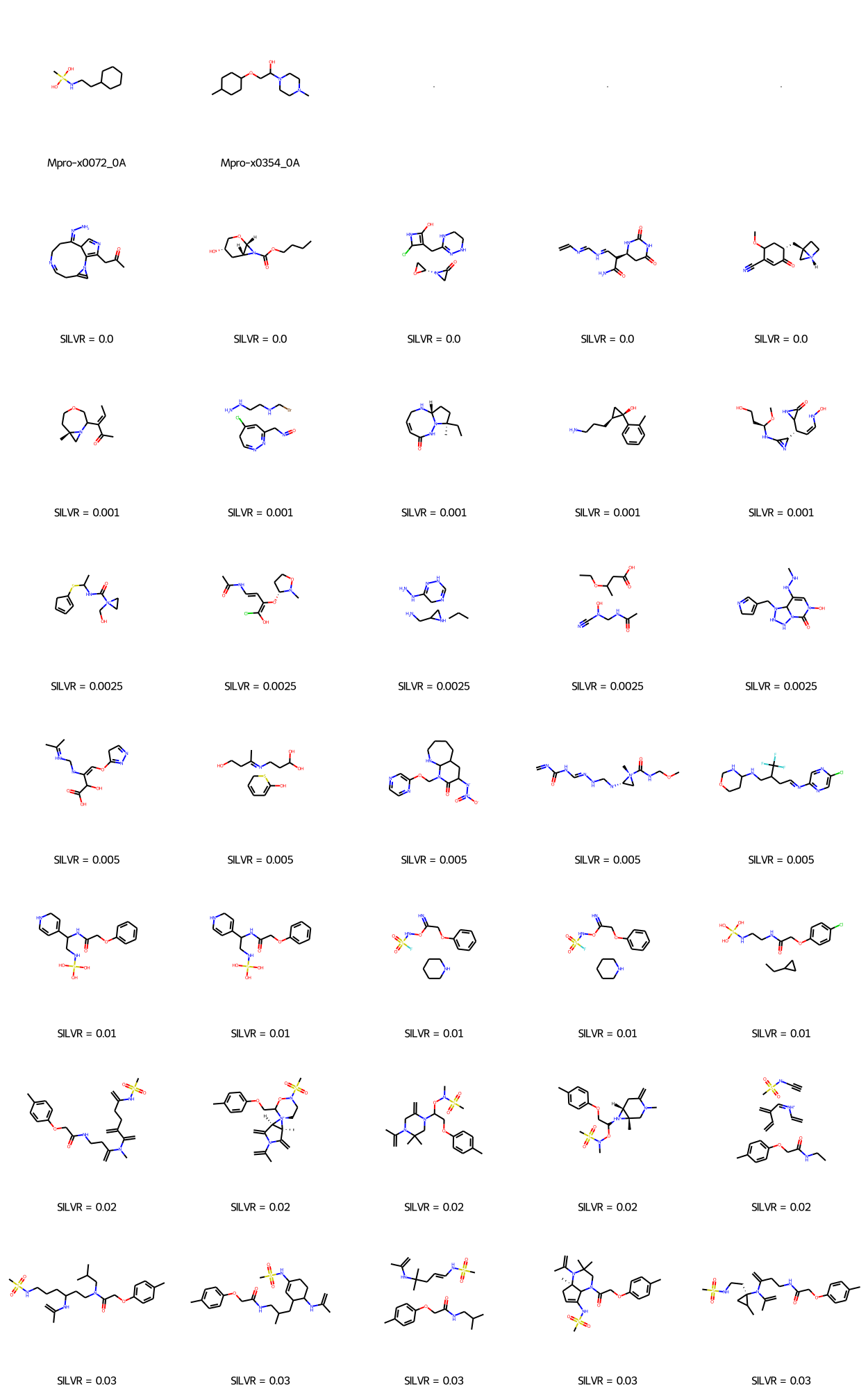}
    \caption{Uncurated samples, showing fragmentation of molecules based on input with two starting structures.}
    \label{fig:uncurated_sample}
\end{figure}

The supporting information consists of a series of figures in addition to the figures in the main paper. Figure~\ref{fig:moonshot_data} contains the 2D structures of all fragments used in the experiments as references. In Figure~\ref{fig:examples_curated} an example of two curated experiments is shown with different SILVR rates $r_{\mathrm{s}}$. Next, Figure~\ref{fig:fragmentation}, shows on average how many molecules were not-fragmented in a set of 1000 samples for different SILVR rates. Figure~\ref{fig:SA_QED} includes synthetic accessibility(~\cite{ertl2009estimation}) and QED~\cite{bickerton2012quantifying, wildman1999prediction}. An example of how far atoms in the reference structure are displaced in the denoising-diffusion process is shown in an example in Figure~\ref{fig:atom_displacement} \textbf{A} for a whole molecule generated and in \textbf{B} for the heterocycle. A summary of experiments around the number of dummy atoms used in linker design experiments is seen in Figure~\ref{fig:dummy_atoms} for $r_{\mathrm{s}}=0.005$ in A and $r_{\mathrm{s}}=0.01$ in B.  The final Figure~\ref{fig:uncurated_sample} shows a series of 2D samples from an uncurated list of samples, clearly showing examples of fragmented molecules.

\section{Modified SILVR with dummy atoms}
The SILVR protocol guides latent atoms according to a one-to-one mapping with reference atoms. This model is limited to reference molecules that somewhat overlap. Two fragments that are sufficiently far apart will not be successfully linked together. A universal EDM protocol for interpreting fragment data would ideally be able to do both fragment merging and linking.

Dummy atoms are latent space atoms that do not have a mapping to a reference atom, and instead are free to explore the latent space unguided. It was hoped that in the case of disconnected fragments, these dummy atoms would be able to form a linker. In practice, linker formation was observed both with and without dummy atoms. It was also observed that dummy atoms often populate unfilled valences as hydrogen atoms. 

The total number of atoms used by the model was set as the sum of reference ($n_r$) and dummy atoms ($n_d$). The SILVR vector was created such that the first $n_r$ indices held the value of $r_{\mathrm{S}}$ defined in the experiment protocol, and the subsequent $n_d$ indices were set to zero so that SILVR would not be applied to the dummy atoms. 

The centre of geometry for the reference coordinates was initially aligned at zero. Within the denoising loop, the EDM sampled new coordinates for all atoms. The new coordinates for dummy atoms were then added to the reference coordinates, and the modified reference coordinates were re-aligned at zero. The SILVR equation was applied as described in the main paper, only refining the atoms mapped to a reference. Through each iteration of denoising, the coordinates of the dummy atoms within the reference were continually updated, and the centre of geometry of these coordinates was re-aligned at zero. The total shift in the centre of geometry of the reference was tracked and subtracted from the final sampled molecule.

\newpage

\newpage

\end{appendix}

\end{document}